\documentclass[runningheads]{llncs}
\usepackage[T1]{fontenc}
\usepackage{graphicx}
\usepackage{verbatim}
\usepackage{xcolor}
\usepackage{booktabs}
\usepackage{tabularx}
\usepackage{multirow}

\begin{document}
\title{Productivity and Collaboration in Hybrid Agile Teams: An Interview Study}
%
%
\author{Elisabeth Mo\inst{1,2} \and
Jefferson Seide Molléri\inst{2}\orcidID{0000-0001-5629-5256} \and
Asle Fagerstrøm\inst{2}\orcidID{0000-0002-8854-1658}}
\authorrunning{Mo, Molléri and Fagerstrøm}
%
\institute{DNB Bank ASA, Olso, Norway\\
\url{http://www.dnb.no/}\\
\email{elisabeth\_mo@live.no} \and
Kristiania University of Applied Sciences, Olso, Norway\\
\url{http://www.kristiania.no}\\
\email{\{asle.fagerstrom,jefferson.molleri\}@kristiania.no}}
\maketitle              
\begin{abstract}

Hybrid work has become a reality post-pandemic, transforming how Agile teams deliver value, collaborate, and adapt. This study investigate how hybrid settings influence productivity and collaboration through nine interviews with three Norwegian Agile teams. Our findings show that hybrid work reduces informal interaction, creates uneven participation, and increases reliance on digital tools. Agile ceremonies became alignment anchors, while trust, communication, and tool support mediate team effectiveness. Hybrid Agile work is an evolving field that requires tailored structures to support inclusion, team cohesion, and sustainable performance.

\end{abstract}

\section{Introduction}

Hybrid work has become a feature of the modern workplace, with many co-located teams now operating in between digital and physical spaces. Several years post-pandemic, organizations embraced flexibility and autonomy, yet teams continue to struggle with alignment and cohesion across distance~\cite{Moe2021,Molleri2024,Souza2022}. These tensions are particularly visible in Agile environments, which emphasize close collaboration, knowledge sharing, and effective task coordination~\cite{Manifesto2001,Dingsoyr2012}.

Hybrid settings disrupt the conditions under which Agile teams work due to scheduling conflicts, miscommunication, and a weakened sense of team cohesion~\cite{Behutiye2024,Hanzis2024}. Beyond that, it raises concerns related to sustainable pace, psychological well-being, and equal participation in decision-making~\cite{Steegh2025,Hanzis2024}.

This study explores these challenges through a qualitative investigation of three Agile teams from Norwegian public and private organizations. We address the following research question: \textit{How do hybrid work environments influence productivity and collaboration within Agile teams?} Our contribution is an empirical account of how Agile practices are being adapted to sustain inclusion, well-being, and effectiveness in hybrid settings.

\section{Background}

Agile methods were originally built around face-to-face interaction, short feedback loops, and collective ownership of work~\cite{Manifesto2001,Dingsoyr2012}. In hybrid work, these foundations are increasingly strained. Informal communication and spontaneous coordination become harder to sustain~\cite{Ly2025}, while digital tools intended to bridge distance can introduce friction when used inconsistently, leading to fragmented awareness and slower decisions~\cite{Behutiye2024}.

Hybrid work blends remote flexibility with periodic office presence~\cite{Hanzis2024}, offering autonomy and work–life integration but complicating Agile's reliance on continuous collaboration. Remote members may feel excluded when decisions occur among co-located colleagues~\cite{Molleri2024}, reinforcing the need for equal access to information, and support for diverse working rhythms~\cite{Kennedyd2024}. Hybrid settings expand talent pools and enable flexible scheduling, but require deliberate coordination structures to maintain team cohesion and ensure inclusion~\cite{Adzgauskaite2025,Souza2022,Moe2021}.

Hybrid conditions directly affect \textit{productivity} (how efficiently teams deliver value) and \textit{collaboration} (their ability to share knowledge, coordinate work, and maintain alignment across locations). Reduced communication bandwidth limits tacit knowledge sharing and non-verbal cues that foster trust and shared understanding~\cite{Adzgauskaite2025}. Participation in Agile ceremonies can become uneven, with remote members struggling to contribute to discussions dominated by those on-site~\cite{Hanzis2024,Molleri2024}. Although tools such as Slack, Teams, and Jira are essential for distributed coordination, inconsistent usage can delay feedback and disrupt workflow~\cite{Ly2025,Behutiye2024}.

\section{Method}

\textbf{Data collection.} Data were collected in 2025, when hybrid work practices had stabilized post-pandemic. The study involved three Agile teams in Norway: two from the financial sector (DNB - \textit{Den Norske Bank}) and one from the public sector (\textit{Digitaliseringsdirektoratet}). 

Participants included software engineers, developers, product owners, and managers, ranging from early-career to experienced (10+ years) professionals. Teams varied in size, roles, and Agile maturity, and worked in cross-functional setups. All teams operated in partially co-located arrangements with regular office days used for Agile ceremonies (such as sprint planning), and collaborative sessions, while development work largely occurred remotely. This setting reflects common hybrid Agile practices in contemporary organizations.

We conducted nine semi-structured interviews with practitioners via Microsoft Teams in March 2025. Interviews lasted 45–70 minutes, and were recorded with participant consent. To ensure confidentiality, all statements were anonymized. Participants received the interview guide in advance.

The IMOI framework allowed us to explore how team dynamics unfold through \textit{inputs} (e.g., hybrid work conditions, tool use), \textit{mediators} (communication quality, shared awareness), \textit{outcomes} (productivity and coordination effectiveness), and \textit{impacts} (adaptability, sustainability, and well-being). These codes were then grouped into higher-level themes, which form the basis of the results in Section~\ref{sec:results}.

\begin{table*}[h]
\vspace{-1em}
\centering
\small
\begin{tabularx}{\textwidth}{X l}
\toprule
\textbf{Interview Quote} & \textbf{Assigned Codes} \\
\midrule

\multirow{5}{=}{\textit{``When we're all working remotely, it's harder to read the room during planning meetings. Sometimes, I can't tell if the team actually agrees or if they're just quiet''.}}
& - \textit{Input:} remote communication challenges\\
& - \textit{Input:} lack of non-verbal cues\\
& - \textit{Mediator:} unclear team alignment \\
\\
\\
\midrule

\multirow{4}{=}{\textit{``Some developers prefer working late at night, while others during the day. That can sometimes delay feedback and reviews''.}}
& - \textit{Input:} timing misalignment\\
& - \textit{Input:} asynchronous work challenges\\
& - \textit{Mediator:} impact on feedback \\
\\
\midrule

\multirow{4}{=}{\textit{``Having one or two set days in the office makes a difference. We use that time for workshops or tougher decisions that are hard to do efficiently online''.}} 
& - \textit{Input:} hybrid schedule structure\\
& - \textit{Mediator:} office time for high-impact tasks\\
& - \textit{Outcome:} in-person decision-making \\
\\
\bottomrule
\end{tabularx}
\caption{Interview quotes and their corresponding assigned codes.}
\label{tab:quotes_codes}
\end{table*}



\section{Results}
\label{sec:results}

\subsection{Collaboration in Hybrid Agile}

Participants emphasized that hybrid work has transformed how Agile teams collaborate. Without the spontaneous interactions of co-located work, teams depend more on structured communication, leading to gaps in shared awareness and uneven participation. One participant explained: \textit{``Sometimes, I just hear about a small decision in passing, and it turns out it was made while I was working from home''.} Collaboration challenges reflect \textit{mediators} such as weakened team processes that are influenced by hybrid \textit{inputs} like asynchronous availability. The main findings in this theme are:

\begin{itemize}
    \item Hybrid setups created gaps in information flow when on-site discussions were not shared digitally.
    \item Office-based decisions may exclude remote members, causing misalignment.
    \item Fewer informal interactions limited clarifications and relationship-building.
    \item Effective collaboration depended on team culture, including trust, openness, and proactive communication.
    \item Agile ceremonies served as alignment anchors, compensating for the loss of informal coordination.
    \item Retrospectives are safe spaces for honest feedback and addressing tensions.
\end{itemize}

\subsection{Communication and Tool Use}

Digital tools emerged as \textit{inputs}, both enablers and constraints for hybrid Agile work. Teams rely on Slack, Microsoft Teams, and Jira to sustain transparency and asynchronous coordination, yet inconsistent use and unclear conventions frequently hinder efficiency. As one interviewee pointed out: \textit{``We often assume something has been communicated, but in reality, it depends on who was in the room or which channel it was posted in''}. Communication tools are \textit{inputs} that influence \textit{mediators} such as coordination and shared understanding.

\begin{itemize}
    \item Digital tools help bridge distance and support asynchronous coordination.
    \item Async updates supported flexibility and protected focus time.
    \item Inconsistent communication routines led to missed updates and uneven engagement within teams.
    \item Misalignment in communication expectations created confusion about urgency, intent, and decision status.
    \item Remote collaboration made work handoff more difficult, reducing shared understanding between roles, e.g. designers and developers.
\end{itemize}






\subsection{Adaptability and Agile Mindset}

Adaptability emerged as a defining \textit{impact} for teams that thrived in hybrid settings. Yet, this adaptability was uneven across members and often shaped by organizational culture. It reflects a key \textit{mediator}, that mean, an emergent team process through which hybrid work \textit{inputs} are managed. An interviewee noted: \textit{``We spend a lot of time in meetings. Sometimes I wish we had more blocks of quiet time to actually do the work.''}  The main codes identified are:

\begin{itemize}
    \item Sprint planning provides direction, but teams often adjust priorities mid-sprint to accommodate shifting availability in hybrid work.
    \item Variation in Agile maturity affected coordination, with some members applying Agile principles more consistently than others.
    \item Greater autonomy enable teams to adapt more effectively to hybrid by re-aligning work without waiting for managerial decisions.
\end{itemize}

\subsection{Productivity and Performance Drivers}

Participants linked productivity to disciplined Agile practice and mindset. The hybrid setting introduced both enablers and barriers to maintaining flow. One participant explained: \textit{``We're most effective when we're trusted to make the call and adjust as needed.''} The productivity \textit{outcomes} described here emerge from earlier relationships between \textit{inputs and mediators} such as work arrangement → coordination and digital tools → communication. The main codes identified are:

\begin{itemize}
    \item Sprint planning helped hybrid teams set shared goals and identify work dependencies.
    \item Frequent digital meetings interrupted work needed for development tasks.
    \item Dependencies on remote actors from outside the Agile team caused delivery delays.
\end{itemize}

\subsection{Work–Life Balance and Satisfaction}

Hybrid work also reshaped how participants experienced motivation, well-being, and sustainable performance. These effects represent an \textit{impact} with broader consequences for individuals beyond immediate team \textit{outcomes}, and they may cycle back to influence future \textit{inputs} such as availability and participation. As noted by a participant: \textit{``We're technically allowed to work from home, but it's kind of expected that you show up a few days a week – no one says it, but it's there.''} The main topics discussed are:

\begin{itemize}
    \item Clear work–life boundaries were respected in some teams, supporting balance and reducing burnout risk.
    \item Unclear expectations to be physically present create subtle pressure despite hybrid policies.
    \item Agile practices, such as short iterations and limiting work-in-progress, support a sustainable pace under increased coordination demands.
    \item Learning became more self-driven in hybrid settings, as opportunities for informal knowledge transfer decreased.
\end{itemize}

\section{Discussion}

Our findings show that Agile practices are not simply transferred into hybrid settings but adapted through ongoing experimentation. Teams reconfigure ceremonies and communication norms to sustain collaboration and productivity across physical and digital spaces.

Interpreting the findings through the IMOI framework~\cite{Ilgen2005} shows how hybrid work reshapes core \textit{inputs} (communication channels, routines, and degrees of co-location) creating both flexibility and new sources of friction. While digital tools support visibility and asynchronous work, decisions made informally in the office can unintentionally exclude remote members, leading to misalignment.

These inputs also shaped \textit{mediators} such as communication quality, trust, and cohesion. Agile ceremonies served as crucial alignment mechanisms that helped compensate for reduced informal contact, yet their effectiveness depended heavily on team culture. Teams with strong norms of openness and proactive communication navigated hybrid constraints more effectively than teams with inconsistent tool use or unclear expectations.

The effects appeared in team \textit{outcomes} and longer-term \textit{impacts}. Productivity and adaptability remained high when teams maintained explicit communication norms and used ceremonies intentionally, but were hindered by external dependencies, meeting overload, or uneven Agile maturity. Sustainable hybrid work required balancing autonomy with inclusive decision-making and clear work–life boundaries. Teams that practiced transparency and respected personal time reported greater engagement.

\subsection{Practical Implications}

Our findings support actionable recommendations for strengthen hybrid Agile:

\begin{itemize}
    \item \textit{Establish intentional communication policies:} define how decisions are shared, how tools are used, and how remote team members are included in informal discussions.
    
    \item \textit{Promote decentralized decision-making:} allow teams to act autonomously while ensuring alignment through shared goals and artifacts.

    \item \textit{Balance ceremonies with focus time: } limit synchronous meetings to essential coordination and use asynchronous updates to preserve deep work.

    \item \textit{Invest in trust-building initiatives:} use retrospectives and team check-ins to express concerns, celebrate progress, and reinforce psychological safety.

    \item \textit{Support sustainable pacing:} encourage boundaries between work and personal life and ensure leaders to be exemplars of these behaviors.
\end{itemize}

These recommendations apply to both team leaders and organizations. Leaders can use them to frame discussions on shared norms rather than impose top-down rules, while organizations must ensure inclusive and flexible participation. Hybrid agility succeeds when individuals, teams, and structures evolve together within a coherent ecosystem.

\subsection{Comparison with Related Work}

Our findings align with existing research on hybrid Agile work while offering a more mature perspective compared to early pandemic studies (e.g.,~\cite{Neumann2022,Tran2022}). By 2025, hybrid work had become a stable and accepted model, and participants described persistent practices (such as hybrid schedules and implicit presence expectations) rather than temporary adaptations.

Consistent with prior work~\cite{Molleri2024,Neumann2022,Tran2022}, we found that hybrid settings reduce informal interactions and create risks of excluding remote members. Our results extend this by clarifying how such gaps arise, for example through micro-decisions or ad-hoc clarifications occurring in the office. Literature emphasizes the need for deliberate communication structures~\cite{Molleri2024}; our study reinforces this and shows that Agile ceremonies act as alignment anchors, though their effectiveness depends strongly on trust, openness, and shared norms.

We further contribute by highlighting how inclusion and trust mediate the link between communication quality and productivity - dimensions underexplored in previous studies. We also identify challenges such as ceremony overload, external dependencies, and uneven Agile maturity, which influence hybrid effectiveness but are rarely discussed. Finally, our findings extend the notion of sustainability beyond job satisfaction~\cite{Tran2022,Kennedyd2024}, showing how hybrid work amplifies Agile's ``sustainable pace'' principle by shaping long-term team cohesion and well-being.

\section{Limitations and Threats to validity} 

The main limitations of our study (according to \cite{Merriam2016}) are:

\textit{Credibility} may be affected by the use of self-reported experiences and researcher interpretation. We mitigated this through triangulation across teams, and iterative checks against the IMOI framework.

\textit{Transferability} is strongest for mature hybrid settings, as data were collected after hybrid work had stabilized. While convenience sampling may limit breadth, we included participants with varied roles and experience levels.

\textit{Dependability} was supported by using a semi-structured interview guide, applying consistent procedures across interviews, and documenting coding steps. Categories were refined iteratively to ensure a traceable analytic process.

\textit{Confirmability} was strengthened through reflexive practices, cross-participant validation, and the IMOI framework. The primary researcher’s employment in one case organization remains a potential source of bias.

Regarding \textit{ethical considerations}, participants were informed of the study’s purpose, anonymity, and withdrawal rights. Data handling followed established qualitative research guidelines~\cite{Flick2018}.

Finally, our findings are \textit{contextually limited} by a small sample and a single snapshot of hybrid Agile practice. Future research could extend this work through longitudinal or cross-cultural studies or by integrating behavioral or performance data.

\section{Conclusion}

Hybrid work has transformed how Agile teams collaborate, communicate, and maintain shared understanding. Although it offers flexibility and autonomy, it also introduces risks such as misalignment, uneven participation, and reduced access to informal interactions that are core to Agile processes. Our study shows that hybrid settings reinterpret Agile practices to fit new conditions. 
We also found out that:

\begin{itemize}
\item Hybrid Agile works best when treated as an evolving process.
\item It requires new practices rather than direct reuse of co-located routines.
\item Effective hybrid work balances synchronous and asynchronous collaboration.
\item Cohesion depends on clear communication and equitable participation.
\item Psychological safety, trust, and shared ownership remain essential values.
\end{itemize}

Future work could extend these findings through longitudinal or cross-context studies. Quantitative or behavioral data would complement our qualitative insights by revealing how productivity and collaboration unfold in practice, further clarifying how Agile principles transform in hybrid work environments.

%
%
%
\bibliographystyle{splncs04}
\bibliography{_references}

@article{Adzgauskaite2025,
  author    = {Adzgauskaite, M. and Tam, C. and Martins, R.},
  title     = {What helps Agile remote teams to be successful in developing software? Empirical evidence},
  journal   = {Information and Software Technology},
  volume    = {177},
  year      = {2025},
  pages     = {107593},
  doi       = {10.1016/j.infsof.2024.107593}
}

@misc{Manifesto2001,
  author    = {{Agile Manifesto}},
  title     = {Manifesto for Agile Software Development},
  year      = {2001},
  howpublished = {\url{http://agilemanifesto.org/}}
}

@article{Behutiye2024,
  author    = {Behutiye, N. W. and Tripathi, N. and Isomursu, M.},
  title     = {Adopting Scrum in Hybrid Settings, in a University Course Project: Reflections and Recommendations},
  journal   = {IEEE Access},
  volume    = {12},
  pages     = {105633--105650},
  year      = {2024},
  doi       = {10.1109/ACCESS.2024.3434662}
}

@article{Ly2025,
  title={The power of words in Agile vs. Waterfall development: Written communication in hybrid software teams},
  author={Ly, Delina and Overeem, Michiel and Brinkkemper, Sjaak and Dalpiaz, Fabiano},
  journal={Journal of Systems and Software},
  volume={219},
  pages={112243},
  year={2025},
  publisher={Elsevier}
}

@article{Cooper2016,
  author    = {Cooper, R.},
  title     = {Agile–Stage-Gate Hybrids},
  journal   = {Research Technology Management},
  volume    = {59},
  number    = {1},
  pages     = {21--29},
  year      = {2016},
  doi       = {10.1080/08956308.2016.1117317}
}

@article{Dingsoyr2012,
  author    = {Dingsøyr, T. and Nerur, S. and Balijepally, V. and Moe, N. B.},
  title     = {A decade of agile methodologies: towards explaining agile software development},
  journal   = {Journal of Systems and Software},
  volume    = {85},
  pages     = {1213--1221},
  year      = {2012},
  doi       = {10.1016/j.jss.2015.05.008}
}

@book{Flick2018,
  author    = {Flick, U.},
  title     = {Designing Qualitative Research},
  publisher = {SAGE Publications Ltd},
  year      = {2018},
  doi       = {10.4135/9781529622737}
}

@article{Hanzis2024,
  author    = {Hanzis, A. and Hallo, L.},
  title     = {The Experiences and Views of Employees on Hybrid Ways of Working},
  journal   = {Administrative Sciences},
  volume    = {14},
  pages     = {263},
  year      = {2024},
  doi       = {10.3390/admsci14100263}
}

@article{Ilgen2005,
  author    = {Ilgen, D. R. and Hollenbeck, J. R. and Johnson, M. and Jundt, D.},
  title     = {Teams in organizations: From input-process-output models to IMOI models},
  journal   = {Annual Review of Psychology},
  volume    = {56},
  pages     = {517--543},
  year      = {2005},
  doi       = {10.1146/annurev.psych.56.091103.070250}
}

@article{Kennedyd2024,
  title={Agile practices and it development team well-being: Unveiling the path to successful project delivery},
  author={Kennedyd, I Sarmann and Zadeh, Adel A and Choi, Jeonghwan and Alborz, Shawn},
  journal={Engineering Management Journal},
  pages={1--13},
  year={2024},
  publisher={Taylor \& Francis}
}

@book{Merriam2016,
  author    = {Merriam, Sharan B. and Tisdell, Elizabeth J.},
  title     = {Qualitative Research: A Guide to Design and Implementation},
  edition   = {4},
  publisher = {Jossey-Bass: A Wiley Brand},
  year      = {2016},
  isbn      = {978-1-119-00361-8},
  pages     = {346}
}

@article{Moe2021,
  title={Finding the sweet spot for organizational control and team autonomy in large-scale agile software development},
  author={Moe, Nils Brede and {\v{S}}mite, Darja and Paasivaara, Maria and Lassenius, Casper},
  journal={Empirical Software Engineering},
  volume={26},
  number={5},
  pages={101},
  year={2021},
  publisher={Springer}
}

@article{Molleri2024,
  author={Molléri, Jefferson Seide and Mohagheghi, Parastoo},
  journal={IEEE Software}, 
  title={Transformation to a Hybrid Workplace: A Case From the Norwegian Public Sector}, 
  year={2024},
  volume={41},
  number={5},
  pages={70-77},
  doi={10.1109/MS.2024.3368564}
}

@inproceedings{neumann2022,
  title={What remains from COVID-19? Agile software development in hybrid work organization: A single case study},
  author={Neumann, Michael and Habibpour, Daryosch and Eichhorn, Dennis and John, A and Steinmann, Stefan and Farajian, L and M{\"o}tefindt, David},
  booktitle={2022 10th International Conference in Software Engineering Research and Innovation (CONISOFT)},
  pages={29--38},
  year={2022},
  organization={IEEE}
}

@inproceedings{Souza2022,
  title={A grounded theory of coordination in remote-first and hybrid software teams},
  author={de Souza Santos, Ronnie E and Ralph, Paul},
  booktitle={Proceedings of the 44th International Conference on Software Engineering},
  pages={25--35},
  year={2022}
}

@article{Steegh2025,
  title={Understanding how agile teams reach effectiveness: A systematic literature review to take stock and look forward},
  author={Steegh, R and Van De Voorde, K and Paauwe, J},
  journal={Human Resource Management Review},
  volume={35},
  number={1},
  pages={101056},
  year={2025},
  publisher={Elsevier}
}

@mastersthesis{Tran2022,
  title={The Impact of Hybrid Work on Productivity: Understanding the Future of Work: A case study in agile software development teams},
  author={Tran, Lisa},
  year={2022},
  type= {Master’s thesis},
  url= {https://www.diva-portal.org/smash/record.jsf?pid=diva2:1696127}
}

\end{document}